\documentclass[review]{elsarticle}

\usepackage{lineno,hyperref}
\usepackage{amsmath}
\usepackage{amssymb}
\usepackage{algorithmic}
\usepackage{algorithm}
\usepackage{caption}
\usepackage{mathtools}
\usepackage{changes}
\usepackage{multirow}
\usepackage{verbatim}
\usepackage{mathrsfs}
\usepackage{graphicx}
\usepackage{amsmath,amsthm,bm,mathrsfs}

\theoremstyle{plain}

\theoremstyle{definition}

\theoremstyle{remark}

\modulolinenumbers[5]

\journal{ArXiv.org}

\begin{document}

\begin{frontmatter}

\title{Non-intrusive Data-driven ADI-based Low-rank Balanced Truncation}

\author[uz]{Umair~Zulfiqar\corref{mycorrespondingauthor}}
\cortext[mycorrespondingauthor]{Corresponding author}
\ead{umairzulfiqar@shu.edu.cn}
\address[uz]{School of Electronic Information and Electrical Engineering, Yangtze University, Jingzhou, Hubei, 434023, China}
\begin{abstract}
In this short note, a non-intrusive data-driven formulation of ADI-based low-rank balanced truncation is provided. The proposed algorithm only requires transfer function samples at the mirror images of ADI shifts. If some shifts are used in both approximating the controllability Gramian and the observability Gramian, then samples of the transfer function's derivative at these shifts are also needed to enforce Hermite interpolation in the Loewner framework. It is noted that ADI-based low-rank balanced truncation can be viewed as a two-step process. The first step involves constructing an interpolant of the original model at the mirror images of the ADI shifts, which can be done non-intrusively within the Loewner framework. The second step involves reducing this interpolant using low-rank factors of Gramians associated with the interpolation data through the balanced square-root algorithm. This second step does not require any system information, making the overall process non-intrusive with the only required information being samples of the transfer function and/or its derivative at the mirror images of ADI shifts. Furthermore, it is shown that when the order of the reduced model in ADI-based low-rank balanced truncation is selected to match the numerical rank of the low-rank factors of the Gramians, it effectively reduces to standard interpolation at the mirror images of the ADI shift. An illustrative example is provided to explain the proposed approach.
\end{abstract}


\end{frontmatter}

\section{Preliminaries}
Consider an \( n^{th} \)-order linear time-invariant (LTI) system \( G(s) \) represented by the state-space realization  
\[
G(s) = C(sE - A)^{-1}B,
\]  
where \( E \in \mathbb{R}^{n \times n} \), \( A \in \mathbb{R}^{n \times n} \), \( B \in \mathbb{R}^{n \times m} \), and \( C \in \mathbb{R}^{p \times n} \).

Suppose the \( r^{th} \)-order reduced-order model (ROM) \( G_r(s) \) is given by the state-space realization  
\[
G_r(s) = C_r(sE_r - A_r)^{-1}B_r,
\]  
where \( E_r \in \mathbb{R}^{r \times r} \), \( A_r \in \mathbb{R}^{r \times r} \), \( B_r \in \mathbb{R}^{r \times m} \), and \( C_r \in \mathbb{R}^{r \times n} \).

The ROM is derived from \( G(s) \) using Petrov-Galerkin projection, defined as  
\[
E_r = W^T E V, \quad A_r = W^T A V, \quad B_r = W^T B, \quad C_r = C V,
\]  
where \( W \in \mathbb{R}^{n \times r} \), \( V \in \mathbb{R}^{n \times r} \), and both \( V \) and \( W \) are full column rank matrices.

For the remainder of our discussion, we will assume, without loss of generality, that \( V \), \( W \), \( E_r \), \( A_r \), \( B_r \), and \( C_r \) are complex matrices.
\subsection{Interpolatory Loewner framework \cite{mayo2007framework}}
In the Loewner framework, the matrices of the ROM are constructed from frequency domain data as follows:
\[
E_r(i,j) = -\frac{c_i\big(G(\sigma_j) - G(\mu_i)\big)b_j}{\sigma_j - \mu_i}, \quad A_r(i,j) = -\frac{c_i\big(\sigma_j G(\sigma_j) - \mu_i G(\mu_i)\big)b_j}{\sigma_j - \mu_i},
\]
\[
B_r(i,:) = c_i G(\mu_i), \quad C_r(:,j) = G(\sigma_j) b_j,
\]
These matrices satisfy the following interpolation conditions:
\[
G(\sigma_j) b_j = G_r(\sigma_j) b_j, \quad c_i G(\mu_i) = c_i G_r(\mu_i),
\]
for \( i = 1, \dots, r \) and \( j = 1, \dots, r \). When \( \sigma_j = \mu_i \), the expressions simplify to:
\[
E_r(i,j) = -c_i \big(G^\prime(\sigma_j)\big) b_j, \quad A_r(i,j) = -c_i \big(G(\sigma_j) + \sigma_j G^\prime(\sigma_j)\big) b_j,
\]
which satisfy the Hermite interpolation conditions:
\[
c_i G^\prime(\sigma_j) b_j = c_i G_r^\prime(\sigma_j) b_j.
\]
The matrices \( E_r \) and \( A_r \) exhibit a special structure known as the Loewner matrix and shifted Loewner matrix, respectively. This structure gives rise to the name \textit{Interpolatory Loewner framework}.
\subsection{Balanced Truncation \cite{moore1981principal}}
Let \( P \) and \( Q \) denote the controllability and observability Gramians, respectively, defined by the following integral expressions:
\begin{align}
P = \frac{1}{2\pi} \int_{-\infty}^{\infty} (j\omega E - A)^{-1} BB^T (-j\omega E^T - A^T)^{-1} \, d\omega, \label{int1}\\
Q = \frac{1}{2\pi} \int_{-\infty}^{\infty} (-j\omega E^T - A^T)^{-1} C^T C (j\omega E - A)^{-1} \, d\omega. \label{int2}
\end{align}
The Gramians \( P \) and \( Q \) can also be computed by solving the following Lyapunov equations:
\[
APE^T + EPA^T + BB^T = 0,
\]
\[
A^TQE + E^TQA + C^T C = 0.
\]
Next, we compute the Cholesky factorizations of \( P \) and \( Q \) as:
\[
P = Z_p Z_p^T \quad \text{and} \quad Q = Z_q Z_q^T.
\]
The balanced square root algorithm \cite{tombs1987truncated} proceeds as follows. First, compute the singular value decomposition (SVD) of \( Z_q^T E Z_p \):
\[
Z_q^T E Z_p = \begin{bmatrix} U_1 & U_2 \end{bmatrix} \begin{bmatrix} S_1 & 0 \\ 0 & S_2 \end{bmatrix} \begin{bmatrix} V_1^T \\ V_2 \end{bmatrix}.
\]
Finally, the projection matrices \( W \) and \( V \) are constructed as:
\[
W = Z_q U_1 S_1^{-\frac{1}{2}} \quad \text{and} \quad V = Z_p V_1 S_1^{-\frac{1}{2}}.
\]
\subsection{ADI-based Low-rank Balanced Truncation}
As demonstrated in \cite{wolf2016adi}, the ADI method implicitly performs \(\mathcal{H}_2\) pseudo-optimal model order reduction (MOR). In this work, we adopt this equivalent representation of the ADI-based low-rank balanced truncation for a data-driven formulation, rather than relying on the original ADI framework \cite{benner2013efficient}. While our presentation differs slightly from the one in \cite{wolf2016adi}, it remains theoretically equivalent to the original approach.

Assume that all ADI shifts \((\alpha_1, \dots, \alpha_k)\) and \((\beta_1, \dots, \beta_l)\) have negative real parts and that there are no repeated \(\alpha_i\) or \(\beta_i\). This assumption is tied to \(\mathcal{H}_2\)-optimal MOR \cite{gugercin2008h_2}, which requires the ROM to have simple poles. Further, define the matrices \(V \in \mathbb{C}^{n \times km}\) and \(W \in \mathbb{C}^{n \times lp}\) as follows:
\begin{align}
V &= \begin{bmatrix} V_1 & \cdots & V_m \end{bmatrix}, \label{blkV}\\
W &= \begin{bmatrix} W_1 & \cdots & W_p \end{bmatrix}, \label{blkW}
\end{align}
where
\begin{align}
V_j &= \begin{bmatrix} (-\alpha_1 E - A)^{-1} B(:,j) & \cdots & (-\alpha_k E - A)^{-1} B(:,j) \end{bmatrix}, \label{V}\\
W_i &= \begin{bmatrix} (-\beta_1 E^T - A^T)^{-1} C(i,:)^T & \cdots & (-\beta_l E^T - A^T)^{-1} C(i,:)^T \end{bmatrix}, \label{W}
\end{align} for \(j = 1, \dots, m\) and \(i = 1, \dots, p\).

Next, define \(s_\alpha\), \(l_\alpha\), \(s_\beta\), and \(l_\beta\) as:
\[
s_\alpha = \text{diag}(-\alpha_1, \dots, -\alpha_k), \quad s_\beta = \text{diag}(-\beta_1, \dots, -\beta_l),
\]
\[
l_\alpha = \begin{bmatrix} 1 & \cdots & 1 \end{bmatrix} \in \mathbb{R}^{1 \times k}, \quad l_\beta = \begin{bmatrix} 1 & \cdots & 1 \end{bmatrix} \in \mathbb{R}^{1 \times l}.
\]

Let \(p_r^{-1}\) and \(q_r^{-1}\) solve the following Lyapunov equations:
\[
-s_\alpha^* p_r^{-1} - p_r^{-1} s_\alpha + l_\alpha^T l_\alpha = 0,
\]
\[
-s_\beta^* q_r^{-1} - q_r^{-1} s_\beta + l_\beta^T l_\beta = 0.
\]

Decompose \(p_r\) and \(q_r\) into their Cholesky factorizations as \(p_r = z_p z_p^*\) and \(q_r = z_q z_q^*\), respectively. Then, define \(\hat{Z}_p\), \(\hat{Z}_q\), \(\tilde{Z}_p\), and \(\tilde{Z}_q\) as:
\[
\hat{Z}_p = I_m \otimes z_p, \quad \hat{Z}_q = I_p \otimes z_q,
\]
\[
\tilde{Z}_p = V \hat{Z}_p, \quad \tilde{Z}_q = W \hat{Z}_q,
\]
such that \(P \approx \tilde{Z}_p \tilde{Z}_p^*\) and \(Q \approx \tilde{Z}_q \tilde{Z}_q^*\). Finally, the low-rank factors \(\tilde{Z}_p\) and \(\tilde{Z}_q\) can replace the full-rank Cholesky factors \(Z_p\) and \(Z_q\) in the balanced square root algorithm to implement low-rank balanced truncation.
\section{Main Work}
We now present the non-intrusive formulation of ADI-based low-rank balanced truncation. For simplicity, assume that \(km = lp\), meaning both \(V\) and \(W\) have the same column rank. While the Loewner framework permits the definition of rectangular Loewner and shifted Loewner matrices \cite{karachalios2023data}, the assumption \(km = lp\) greatly simplifies the subsequent discussion, and thus, we adopt it here. 

The balanced square root algorithm for ADI-based low-rank balanced truncation proceeds as follows. First, compute the SVD of \(\tilde{Z}_q^T E \tilde{Z}_p\) as:
\[
\tilde{Z}_q^T E \tilde{Z}_p = \hat{Z}_q^T W^T E V \hat{Z}_p = \begin{bmatrix} \hat{U}_1 & \hat{U}_2 \end{bmatrix} \begin{bmatrix} \hat{S}_1 & 0 \\ 0 & \hat{S}_2 \end{bmatrix} \begin{bmatrix} \hat{V}_1^T \\ \hat{V}_2 \end{bmatrix}.
\]

The projection matrices for low-rank balanced truncation are then computed as:
\[
\tilde{W} = W \hat{Z}_q \hat{U}_1 \hat{S}_1^{-\frac{1}{2}}, \quad \tilde{V} = V \hat{Z}_p \hat{V}_1 \hat{S}_1^{-\frac{1}{2}}.
\]

Next, define \(\hat{V}\) and \(\hat{W}\) as:
\[
\hat{V} = \hat{Z}_p \hat{V}_1 \hat{S}_1^{-\frac{1}{2}}, \quad \hat{W} = \hat{Z}_q \hat{U}_1 \hat{S}_1^{-\frac{1}{2}}.
\]
The low-rank truncated balanced ROM \(\tilde{G}_r(s)\) is obtained as:
\[
\tilde{G}_r(s) = \tilde{C}_r (s \tilde{E}_r - \tilde{A}_r)^{-1} \tilde{B}_r,
\]
where the matrices are constructed as:
\begin{align}
\tilde{E}_r &= \hat{W}^* (W^* E V) \hat{V} = I, & \tilde{A}_r &= \hat{W}^* (W^* A V) \hat{V},\nonumber\\
\tilde{B}_r &= \hat{W}^* (W^* B), & \tilde{C}_r& = (C V) \hat{V}.\nonumber
\end{align}
Note that \(V\) and \(W\) in (\ref{blkV}) and (\ref{blkW}), respectively, are block rational Krylov interpolatory projectors that enforce interpolation at \((-\alpha_1, \dots, -\alpha_k)\) and \((-\beta_1, \dots, -\beta_l)\) for all subsystems of \(G(s)\). The system \(G(s)\) can be decomposed into single-input single-output (SISO) subsystems as:
\[
G(s) = \begin{bmatrix} G_{1,1}(s) & \cdots & G_{1,m}(s) \\ \vdots & \ddots & \vdots \\ G_{p,1}(s) & \cdots & G_{p,m}(s) \end{bmatrix}.
\]

Furthermore, \(V_j\) and \(W_i\) in (\ref{V}) and (\ref{W}) are interpolatory projectors that enforce interpolation at \((-\alpha_1, \dots, -\alpha_k)\) and \((-\beta_1, \dots, -\beta_l)\) for individual SISO subsystems \(G_{i,j}(s)\) of \(G(s)\). From standard block interpolation theory, we have:
\begin{align}
E_r&=W^TEV=\begin{bsmallmatrix}E_{r,1,1}&\cdots& E_{r,1,m}\\\vdots&\ddots&\vdots\\E_{r,p,1}&\cdots&E_{r,p,m}\end{bsmallmatrix}=\begin{bsmallmatrix}W_1^TEV_1&\cdots& W_1^TEV_m\\\vdots&\ddots&\vdots\\W_p^TEV_1&\cdots&W_p^TEV_m\end{bsmallmatrix},\nonumber\\
A_r&=W^TAV=\begin{bsmallmatrix}A_{r,1,1}&\cdots& A_{r,1,m}\\\vdots&\ddots&\vdots\\A_{r,p,1}&\cdots&A_{r,p,m}\end{bsmallmatrix}=\begin{bsmallmatrix}W_1^TAV_1&\cdots &W_1^TAV_m\\\vdots&\ddots&\vdots\\W_p^TAV_1&\cdots&W_p^TAV_m\end{bsmallmatrix},\nonumber\\
B_r&=W^TB=\begin{bsmallmatrix}B_{r,1}\\\vdots\\B_{r,p}\end{bsmallmatrix}=\begin{bsmallmatrix}W_1^TB\\ \vdots\\W_p^TB\end{bsmallmatrix},\nonumber\\
C_r&=CV=\begin{bsmallmatrix}C_{r,1}&\cdots&C_{r,m}\end{bsmallmatrix}=\begin{bsmallmatrix}CV_1&\cdots&CV_m\end{bsmallmatrix}\nonumber
\end{align}
where the individual matrices are given by:
\begin{align}
E_{r,p,m}(i,j) &= \frac{G_{p,m}(-\alpha_j) - G_{p,m}(-\beta_i)}{\alpha_j - \beta_i}, \label{my_lv1}\\
A_{r,p,m}(i,j) &= -\frac{\alpha_j G_{p,m}(-\alpha_j) - \beta_i G_{p,m}(-\beta_i)}{\alpha_j - \beta_i},\\
B_{r,p}(i,:) &= \begin{bmatrix} G_{p,1}(-\beta_i) & \cdots & G_{p,m}(-\beta_i) \end{bmatrix},\\
C_{r,m}(:,j) &= \begin{bmatrix} G_{1,m}(-\alpha_j) \\ \vdots \\ G_{p,m}(-\alpha_j) \end{bmatrix}, \label{my_lv4}
\end{align}for \(i = 1, \dots, l\) and \(j = 1, \dots, k\). When \(\alpha_j = \beta_i\), the expressions simplify to:
\begin{align}
E_{r,p,m}(i,j) = -G_{p,m}^\prime(-\alpha_j), \quad A_{r,p,m}(i,j) = \alpha_j G_{p,m}^\prime(-\alpha_j) - G_{p,m}(-\alpha_j).
\end{align}
In summary, these matrices can be constructed non-intrusively using samples of \(G(s)\) at \((-\alpha_1, \dots, -\alpha_k)\) and \((-\beta_1, \dots, -\beta_l)\) within the Loewner framework. If there are common elements in the sets \((-\alpha_1, \dots, -\alpha_k)\) and \((-\beta_1, \dots, -\beta_l)\), samples of \(G^\prime(s)\) at those common points are also required.

Note that \(\hat{Z}_p\) and \(\hat{Z}_q\) do not depend on \(G(s)\), as they are entirely determined by the ADI shifts \((-\alpha_1, \dots, -\alpha_k)\) and \((-\beta_1, \dots, -\beta_l)\). Additionally, \(\hat{V}\) and \(\hat{W}\) can be computed using the Cholesky factors \(\hat{Z}_p\) and \(\hat{Z}_q\) along with the already constructed \(E_r\). In essence, the interim ROM constructed in the Loewner framework can be further reduced using the balanced square root algorithm with the Cholesky factors \(\hat{Z}_p\) and \(\hat{Z}_q\). Thus, the low-rank truncated balanced ROM \(\tilde{G}_r(s)\) can be obtained non-intrusively without accessing the state-space realization \((E, A, B, C)\) during the reduction process.
\subsection{Some Observations}
\begin{enumerate}
  \item When the order \( r \) of the low-rank truncated balanced ROM \(\tilde{G}_r(s)\) is set to \( r = km = lp \), it follows that \(\hat{W}^T = \hat{V}^{-1}\) and \(\tilde{G}_r(s) = G_r(s)\). In this case, the low-rank ADI-based balanced truncation reduces to interpolation at the mirror images of the ADI shifts \((-\alpha_1, \dots, -\alpha_k)\) and \((-\beta_1, \dots, -\beta_l)\). Thus, the low-rank balanced truncation process can be viewed as a two-step procedure. In the first step, \(G(s)\) is reduced using interpolation. Once the order of \(G(s)\) has been sufficiently reduced to alleviate computational complexity, the balanced square root algorithm is applied to the interim ROM to obtain the final ROM of the desired order. Moreover, the success of ADI methods in accurately implementing balanced truncation over the past three decades provides numerical evidence that even a small number of interpolation points can effectively produce a truncated balanced ROM. While there has been some interest in the MOR community in generating truncated balanced ROMs via interpolation \cite{ionescu2012balancing,kawano2023gramian}, it is not duly appreciated in the literature that low-rank Krylov-subspace methods or ADI methods for balanced truncation do produce truncated balanced realizations within the interpolation framework when they are accurate.
  \item In the interpolatory Loewner framework, the singular values of the matrix \(\begin{bmatrix} E_r & A_r \end{bmatrix}\) play a crucial role in determining the final order of the reduced model. Similarly, in low-rank balanced truncation, the singular values of \(\hat{Z}_q^T E_r \hat{Z}_p\) provide insight into the appropriate final order of the ROM \(\tilde{G}(s)\), as this matrix contains approximations of the dominant Hankel singular values of \(G(s)\). Apart from this distinction, all other steps remain the same as in the interpolatory Loewner framework, with the only difference being the use of block interpolation instead of tangential interpolation. This new tool, \(\hat{Z}_q^T E_r \hat{Z}_p\), which can be computed solely from interpolation data (since \(E_r\) is already constructed in the interpolatory Loewner framework), can also be utilized within the interpolation framework to estimate the Hankel singular values. These estimates can then guide the selection of the ROM's order based on the dominant Hankel singular values.
  \item If the integrals in (\ref{int1}) and (\ref{int2}) are approximated using numerical integration with a suitable quadrature rule, it directly yields low-rank factors \( P \approx \tilde{Z}_p \tilde{Z}_p^* \) and \( Q \approx \tilde{Z}_q \tilde{Z}_q^* \), as demonstrated in \cite{imran2015model}. In this approach, the low-rank factors can be computed through numerical integration using a quadrature rule. This strategy was also employed in data-driven quadrature-based balanced truncation \cite{gosea2022data}. Once the quadrature-based low-rank factors are used to approximate \( Z_q^T E Z_p \approx \tilde{Z}_q^T E_r \tilde{Z}_p \), a non-intrusive formulation similar to the one presented in this work is obtained. We note that the essence of the ADI-based method and quadrature-based balanced truncation is fundamentally the same, as explained below. The core idea of numerical integration for a function \( f(x) \) involves first approximating \( f(x) \) by a polynomial \( p(x) \) that interpolates \( f(x) \) at specific points. Instead of integrating \( f(x) \), the polynomial \( p(x) \) is integrated, as the integration of \( p(x) \) is straightforward. This principle underlies most quadrature rules in numerical integration. Interestingly, the ADI method follows a similar strategy. It first computes a rational interpolation of 
\[
F(s) = (s E - A)^{-1} B B^T (s^* E^T - A^T)^{-1}
\]
as 
\[
\tilde{F}(s) = V (s E_r - A_r)^{-1} B_r B_r^T (s^* E_r^T - A_r^T)^{-1} V^*,
\]
which interpolates \( F(s) \) at the mirror images of the ADI shifts \( (-\alpha_1, \dots, -\alpha_k) \). The ADI method then implicitly integrates \( \tilde{F}(s) \) instead of \( F(s) \) to obtain \( P \approx V \hat{Z}_p \hat{Z}_p^* V^T \). Similarly, \( Q \approx W \hat{Z}_q \hat{Z}_q^* W^T \) is computed. The elegance of the ADI method lies in its avoidance of dealing with weights, which are typically required in various quadrature rules.
\end{enumerate}
\section{Illustrative Example}
Consider an \(8^{th}\)-order nonsquare system with three inputs and two outputs, represented by the following state-space realization:
\begin{align}
E &= \begin{bsmallmatrix}
0.1926  &  0.6769  &  0.3832  &  0.2205  &  0.3438  &  0.3201  & -0.1136 &  -0.1047\\
0.6233 &  -0.2939 &  -0.0498 &  -0.1480 &   0.3180   & 0.2294  & -0.0479  &  0.2938\\
0.3627  &  0.3152 &   0.3227 &  -0.1538 &  -0.6426  & -0.3885 &  -0.1011  &  0.1338\\
-0.2109 &  -0.2730  &  0.2535 &   0.6097  & -0.2312  &  0.0855 &   0.1367  & -0.1491\\
-0.1163 &  -0.0004 &  -0.0750 &  -0.5572  & -0.3071  &  0.5018  & -0.1349  & -0.4473\\
-0.0921  &  0.0343  &  0.4109 &  -0.2228  &  0.1765   & 0.0255  &  0.5661  & -0.2781\\
0.3840  &  0.1811 &  -0.5404  &  0.2717  & -0.0085 &  -0.1855   & 0.0754  & -0.6157\\
0.0115  & -0.0568  &  0.0827  & -0.3002   & 0.3121 &  -0.5534  &  0.2686  & -0.1624
\end{bsmallmatrix},\nonumber\\
A &= \begin{bsmallmatrix}
0.1862  & -0.5917 &  -0.1650 &  -0.9115  & -0.7093 &  -0.2857   & 0.3976  & -0.2267\\
-0.1292  &  0.2346  &  0.0235  &  0.2210  &  0.0896  & -0.0648 &  -0.1632  &  0.1099\\
-1.0098  & -0.2598  & -0.1364 &  -0.6800  &  1.6082  &  1.1541  &  0.0975 &  -0.2698\\
-0.3744  & -0.2510  & -0.0262 &  -1.0588  &  0.2506  &  0.2385  &  0.1537 &  -0.1218\\
0.3892 &   0.4394 &   0.3248 &   1.4637  & -0.2898 &  -0.5814 &  -0.3081 &   0.5925\\
0.3352  &  0.4726 &  -0.7287  &  1.3470 &  -0.6593 &  -0.6707 &  -1.6304 &   1.1902\\
-0.4746  & -0.4897 &   0.9502 &  -1.1089 &   0.2394 &  -0.6600 &   0.1590 &   0.7444\\
0.3554  &  0.4451  & -0.8086 &   1.0669 &  -0.2288 &   1.0456  &  0.1129 &  -0.2125
\end{bsmallmatrix},\nonumber\\
B &= \begin{bsmallmatrix}
-1.7673 &  -0.5417  &  1.4269\\
1.2705 &   0.2508  &  0.6108\\
-0.0455 &  -0.2633  &  2.4580\\
0.2551 &   0.3691  & -0.8544\\
0.3367 &  -0.3287  & -0.0206\\
0.5722 &  -0.9936  &  0.7582\\
-0.9008 &   1.3205  & -0.1470\\
0.6285 &  -0.7686  &  0.4623
\end{bsmallmatrix},\nonumber\\
C &= \begin{bsmallmatrix}
1.4466  & -1.0413 &  -0.1511 &  -0.7822 &   1.1399  &  1.0843  &  0.1865 &   0.3566\\
0.4855  &  1.4606  &  0.7186  &  0.9876  &  1.4261 &   0.9412 &  -0.5588 &   0.1842
\end{bsmallmatrix}.\nonumber
\end{align}
The Hankel singular values of this system are: \(24.3760\), \(6.4380\), \(4.6620\), \(0.5519\), \(0.0985\), \(0.0677\), \(0.0309\), \(0.0035\).

Using the ADI shifts \((-2.3710, -1.1434)\) for the controllability Gramian and \((-0.0195, -0.1543, -0.3513)\) for the observability Gramian, the following \(6^{th}\)-rank factors are obtained using the low-rank Cholesky factor-based ADI (LRCF-ADI) method \cite{benner2013efficient}:
\[
\tilde{Z}_p = \begin{bsmallmatrix}
-0.6466 &  -0.2227&   -1.4898  &  0.5659  & -0.0238  &  1.4611\\
1.8516 &   0.3883 &  -1.3579 &  -1.3736  & -0.5432  &  1.4384\\
-0.3853  &  0.9011  & -1.0101  &  0.1167 &  -0.4042 &   0.5705\\
0.1670 &  -0.4723  &  0.7534  &  0.4572 &   0.0921 &  -0.8769\\
0.1037 &   0.0536  &  0.1489  &  0.1199  &  0.3450  &  0.4309\\
-0.0334 &  -0.0300 &   0.0241  &  0.2669 &  -0.2894 &   0.3827\\
-0.0098 &  -0.0708 &  -0.0433  & -0.3417 &   0.1094 &  -0.4951\\
-0.0170 &   0.0287 &  -0.0414  & -0.0351 &   0.0017 &  -0.0638
\end{bsmallmatrix},
\]
\[
\tilde{Z}_q = \begin{bsmallmatrix}
-0.4407 &  -0.9953 &  -0.6352  & -1.6598 &   0.3442 &  -0.6511\\
-3.6806  & -1.3208 &  -2.8997 &  -0.8665 &  -0.4481 &   0.2312\\
0.0930  & -0.3181 &   0.0484 &  -0.4378  &  0.2807  &  0.1940\\
3.2533  &  1.1672 &  -1.0053 &  -0.0037 &  -0.3344 &  -0.1048\\
2.1278  &  0.3444  & -0.7817 &  -0.5453  & -0.0709 &   0.1792\\
0.0447 &  -0.0460 &   0.1297 &  -0.0841 &  -0.0628 &  -0.0457\\
-0.8279 &  -0.2170 &   0.4058 &   0.1195  & -0.0289 &  -0.0589\\
-0.2874 &  -0.0689&    0.1462&    0.0490  & -0.0216&   -0.0260
\end{bsmallmatrix}.
\]
Using the balanced square root algorithm \cite{tombs1987truncated}, the following \(3^{rd}\)-order low-rank truncated balanced ROM is obtained:
\[
A_r = \begin{bsmallmatrix}
-0.0805 &  -0.0068 &  -0.0286\\
-0.0794 &  -0.4975  &  0.0790\\
-0.0604 &  -0.0613  & -0.0301
\end{bsmallmatrix},
\]
\[
B_r = \begin{bsmallmatrix}
0.3069 &   0.0390  &  2.3280\\
-2.0720 &  -0.6416  &  0.5037\\
0.2104 &  -0.8376 &   0.1153
\end{bsmallmatrix},
\]
\[
C_r = \begin{bsmallmatrix}
0.7776 &  -1.5886 &   0.3494\\
1.1531 &   1.4209  &  0.1723
\end{bsmallmatrix}.
\]
The Hankel singular values of this ROM, which closely approximate the three largest Hankel singular values of the original system, are: \(24.5142\), \(7.6744\), and \(4.6724\).

For the proposed data-driven implementation, the Cholesky factors \( z_p \) and \( z_q \) can be computed directly from the ADI shifts as follows:
\[
z_p = \begin{bmatrix} 6.2341 & 0 \\ -4.0565 & 1.5122 \end{bmatrix}, \quad z_q = \begin{bmatrix} 0.2845 & 0 & 0 \\ -1.1603 & 1.4257 & 0 \\ 1.0733 & -1.9813 & 0.8382 \end{bmatrix}.
\]
Using samples of \( G(s) \) at the ADI shifts, the interim interpolatory \( 6^{th}\)-order ROM can be constructed from (\ref{my_lv1})-(\ref{my_lv4}) as:
\begin{align}
E_r &= \begin{bsmallmatrix}
3.2410 & 5.9157 & 2.3298 & 4.6802 & 11.5376 & 23.5839 \\
2.3794 & 4.3492 & 0.6151 & 1.1577 & 3.5783 & 7.3878 \\
1.6852 & 3.0690 & 0.3840 & 0.7034 & 1.7411 & 3.6508 \\
-0.8992 & -1.4181 & -0.7641 & -1.3203 & 8.2411 & 16.1562 \\
-0.8005 & -1.2676 & -0.8065 & -1.4910 & 4.1010 & 7.9563 \\
-0.6781 & -1.0716 & -0.5696 & -1.0573 & 2.5235 & 4.8930
\end{bsmallmatrix},\nonumber\\
A_r &= \begin{bsmallmatrix}
-1.2510 & -2.1715 & -0.1693 & -0.3418 & -0.2364 & -0.6261 \\
-0.9470 & -1.6157 & -0.1198 & -0.2544 & 0.0908 & 0.0539 \\
-0.7221 & -1.2087 & -0.0798 & -0.1860 & 0.1503 & 0.1965 \\
1.0834 & 1.5940 & 0.3218 & 0.6239 & -0.9583 & -2.0250 \\
0.9774 & 1.4260 & 0.2123 & 0.4196 & -0.4862 & -1.1124 \\
0.8627 & 1.2451 & 0.1366 & 0.2782 & -0.2325 & -0.6211
\end{bsmallmatrix},\nonumber\\
B_r &= \begin{bsmallmatrix}
8.9354 & 5.6931 & 27.5919 \\
6.5887 & 1.5782 & 8.3933 \\
4.7178 & 0.9902 & 3.9778 \\
-3.2154 & -2.1335 & 20.4980 \\
-2.8754 & -2.1244 & 10.2096 \\
-2.4704 & -1.4871 & 6.2158
\end{bsmallmatrix},\nonumber\\
C_r& = \begin{bsmallmatrix}
1.3142 & 2.2868 & 0.2147 & 0.4331 & 0.4614 & 1.0860 \\
-1.1009 & -1.6216 & -0.3367 & -0.6497 & 1.1190 & 2.3400
\end{bsmallmatrix}.\nonumber
\end{align}

The final \( 3^{rd}\)-order truncated low-rank balanced ROM \( \tilde{G}_r(s) \) is obtained by applying the balanced square root algorithm to this interim ROM using the Cholesky factors \( z_p \) and \( z_q \), which are computed solely from the ADI shifts:
\[
\tilde{A}_r = \begin{bsmallmatrix}
-0.0805 & 0.0068 & -0.0286 \\
0.0794 & -0.4975 & -0.0790 \\
-0.0604 & 0.0613 & -0.0301
\end{bsmallmatrix},
\]
\[
\tilde{B}_r = \begin{bsmallmatrix}
0.3069 & 0.0390 & 2.3280 \\
2.0720 & 0.6416 & -0.5037 \\
0.2104 & -0.8376 & 0.1153
\end{bsmallmatrix},
\]
\[
\tilde{C}_r = \begin{bsmallmatrix}
0.7776 & 1.5886 & 0.3494 \\
1.1531 & -1.4209 & 0.1723
\end{bsmallmatrix}.
\]
The Hankel singular values and the transfer function of this ROM are identical to those produced by the LRCF-ADI-based balanced truncation, even though the state-space realization is not the same.

\end{document}